\documentclass{article}

\usepackage{epsfig,layout,amsmath}

\begin{document}

\title{      Application of Kolmogorov complexity and universal codes
to  identity testing and nonparametric testing of serial
independence for time series.
 \footnote {Research  was supported  by the joint
project grant "Efficient randomness testing of random and
pseudorandom number generators" of  Royal Society, UK (grant ref:
15995) and Russian Foundation for Basic Research (grant no.
     03-01-00495.)
 } }
\author{ Boris Ryabko$ ^\dag$,  Jaakko Astola $ ^\ddag$,
  Alex Gammerman $ ^\sharp$
\\ \\
$ ^\dag $ {\small Institute of Computational Technology of
Siberian Branch of Russian Academy of Science.
}
\\
$ ^\ddag$ {\small Tampere University of Technology, Finland.}
\\
$ ^\sharp$ {\small Department of Computer Science, Royal Holloway,
University of London.}
 }
\date{}
\maketitle

\thispagestyle{empty}

\begin{abstract}

We show that  Kolmogorov complexity and such its estimators as
universal codes (or data compression methods) can be applied for
hypotheses testing in a framework of classical mathematical
statistics. The methods for identity testing and nonparametric
testing of serial independence for time series are suggested.
\end{abstract}



\section{Introduction.}

The Kolmogorov complexity, or algorithmic entropy, was suggested
in \cite{Ko} and was investigated in numerous papers; see for
review \cite{Vi}. Now this notation plays important role in theory
of algorithms, information theory, artificial intelligence and
many other fields and is closely connected with such deep
theoretical issues as  definition of randomness, logical basis of
probability theory, randomness and complexity (see \cite{
Hu,Vi,M-L,USS,V-V,V-L, Z-L}). In this paper we show that
Kolmogorov complexity can be applied to hypotheses testing in
framework of mathematical statistics. Moreover, we suggest using
universal codes (or methods of data compression), which are
estimations of Kolmogorov complexity, for testing.
%It turns out that some practically used methods of data compression are quite
%efficient, when they are applied  for hypotheses testing.

In this paper  we consider a stationary and ergodic source (or
process), which generates elements from a finite set (or alphabet)
$A$ and two problems of statistical testing. The first problem is
the identity testing, which is described as follows: a hypotheses
$H_0^{id}$ is that the source has a particular distribution $\pi$
and the alternative hypothesis $H_1^{id}$ that the sequence is
generated by a stationary and ergodic source, which differs from
the source under $H_0^{id}$. One particular case where the source
alphabet $A= \{ 0, 1 \} $ and the main hypothesis $H_0^{id}$ is
that a bit sequence is generated by the Bernoulli source with
equal probabilities of 0's and 1's, is applied to the randomness
testing of random number and pseudorandom number generators.

%, which are used for many purposes including cryptographic,
%modelling and simulation applications.

The second problem is a generalization of the problem of
nonparametric testing for independence of time series. More
precisely, we consider two following  hypotheses: $H_0^{ind}$ is
that the source is Markovian, which memory (or connectivity) is
not larger than $m,\: (m \geq 0),$ and the alternative hypothesis
$H_1^{ind}$ that the sequence is generated by a stationary and
ergodic source, which differs from the source under $H_0^{ind}$.
In particular, if $m=0,$  this is the problem of testing for
independence of time series. This problem is well known in
mathematical statistics and there is an extensive literature
dealing with nonparametric independence testing.

In both cases the testing should be based on a sample $x_1 \ldots
x_t$ generated by the source.

We suggest statistical tests for  identity testing and
nonparametric testing of serial independence for time series,
which are based on  Kolmogorov complexity and  such estimates of
it as universal codes. It is important to note that practically
used so-called archivers can be used for suggested testing,
because they can be considered as methods for estimation of
Kolmogorov complexity.

The outline of the paper is as follows. The next part  contains
definitions and necessary  information. The parts three and four
are  devoted to the identity testing and testing of serial
independence, correspondingly.  The fifth part contains results of
experiments, where the suggested method of identity testing is
applied to  pseudorandom number generators. All proofs are given
in Appendix.

\section{Definitions and Preliminaries.}
First we define stochastic processes (or sources of information).
Consider an alphabet $A=\{a_1,\cdots,a_n\}$ with $n\geq2$ letters
and denote by $A^t$ and $A^*$ the set of all words of length $t$
over $A$ and the set of all finite words over $A$, correspondingly
($A^* = \bigcup_{i=1}^\infty A^i$). Let $\mu$ be a source which
generates letters from $A$. Formally, $\mu$ is a probability
distribution on the set of words of infinite length or, more
simply, $\mu=(\mu^t)_{t\geq1}$ is a consistent set of
probabilities over the sets $A^t\,;\;t\geq1$. By $M_\infty(A)$ we
denote the set of all stationary and ergodic sources, which
generate letters from $A.$ Let $M_k(A) \subset M_\infty(A)$ be the
set of Markov sources with memory (or connectivity) $k, \, k\geq
0. $ More precisely, by definition $\mu \in M_k(A)$ if
\begin{multline}\label{ma}
\mu (x_{t+1} = a_{i_1} / x_{t} = a_{i_2}, x_{t-1} = a_{i_3},\,
...\,, x_{t-k+1} = a_{i_{k+1}},\, ... )  \\
  = \mu (x_{t+1} =
a_{i_1} / x_{t}  = a_{i_2}, x_{t-1} = a_{i_3},\, ...\,, x_{t-k+1}
= a_{i_{k+1}})
\end{multline}
for all $t \geq k $ and $a_{i_1},
a_{i_2}, \ldots \, \in A.$ By definition, $M_0(A)$ is the set of
all Bernoulli (or i.i.d.) sources over $A$ and $M^*(A) =
\bigcup_{i=0}^\infty M_i(A)$ is the set of all finite-memory
sources.

Now we define codes and the Kolmogorov complexity. Let $A^\infty$
be the set of all infinite words $x_1x_2 \ldots $ over the
alphabet $A$. A data compression method (or code) $\varphi$ is
defined as a set of mappings $\varphi_n $ such that $\varphi_n :
A^n \rightarrow \{ 0,1 \}^*,\, n= 1,2, \ldots\, $ and for each
pair of different words $x,y \in A^n \:$ $\varphi_n(x) \neq
\varphi_n(y) .$ Informally, it means that the code $\varphi$ can
be applied for compression of each message of any length $n$ over
alphabet $A$ and the message can be decoded if its code is known.
It is also required that each sequence
$\varphi_n(u_1)\varphi_n(u_2) ...\varphi_n(u_r), r \geq 1,$ of
encoded words from the set $A^n, n\geq 1,$ can be uniquely decoded
into $u_1u_2 ...u_r$. Such codes are called uniquely decodable.
For example, let $A=\{a,b\}$, the code $\psi_1(a) = 0, \psi_1(b) =
00, $  obviously, is not uniquely decodable. It is well known
 that if  a code $\varphi$ is uniquely decodable
then  the lengths of the codewords satisfy the following
inequality (Kraft inequality): $ \Sigma_{u \in A^n}\: 2^{-
|\varphi_n (u) |} \leq 1\:,$ see, for ex., \cite{Ga}. (Here and
below $|v|$ is the length of $v$, if $v$ is a word and the number
of elements of $v$ if $v$ is a set.) It will be convenient to
reformulate this property as follows:

\textbf{Claim 1.} Let $\varphi$ be a uniquely decodable code over
an alphabet $A$. Then  for any integer $n$ there exists a measure
$\mu_\varphi$ on $A^n$ such that
\begin{equation}\label{kra}
|\varphi (u)| \geq - \log \mu_\varphi (u)
\end{equation} for any $u$ from $A^n \,.$ (Here and below $ \log \equiv \log_2$\:.)

(Obviously, the claim  is true for the measure $$\mu_\varphi (u) =
2^{- |\varphi (u) |} / \Sigma_{u \in A^n}\: 2^{- |\varphi (u)
|}).$$ In this paper we will use the so-called prefix Kolmogorov
complexity, whose precise definition can be found in  \cite{Hu,
Vi}. Its main  properties can be described as follows. There
exists a uniquely decodable code $\kappa$ such that i) there is an
algorithm  of decoding (i.e. there is a Turing machine, which maps
$\kappa(u)$ to $u$ for any $u \in A^*$) and ii) for any uniquely
decodable code $\psi,$ whose decoding is algorithmically
realizable,  there exists a constant $C_{\psi}$ that
\begin{equation}\label{ko} |\kappa(u)| - |\psi(u)| < C_{\psi} \end{equation} for any $u \in A^*$.
The prefix Kolmogorov complexity $K(u)$ is defined as the length
of $\kappa(u)$:
 $K(u) = |\kappa(u)|.$  The code $\kappa$ is not unique, but the second
 property
means that codelengths of two codes $\kappa_1$ and $\kappa_2$, for
which i) and ii) is true, are equal up to a constant: $|\:
|\kappa_1(u)| - |\kappa_2(u)| \:| < C_{1,2}$ for any word $u$ (and
the constant $C_{1,2}$ does not depend on $u$, see (\ref{ko}).)
So, $K(u)$ is defined up to a constant.

In what follows we call this value  "Kolmogorov complexity" and
 uniquely decodable codes just "codes".

 We can see from  ii) that the code $\kappa$ is
 asymptotically (up to the constant) the best method of data
 compression, but it turns out that there is no algorithm that
 can calculate the codeword $\kappa(u)$ (and even $K(u)$). That is
 why the code $\kappa$ (and Kolmogorov complexity) cannot be used
 for practical data compression directly. On the other hand,
 so-called universal codes can be realized and, in a certain sense,
 can be used instead of the optimal code $\kappa$, if they are applied for
 compression of sequences generated by any stationary and ergodic
 source. For their description we recall that (as
it is known in Information Theory)  sequences $x_1 ... x_t,$
generated by a  source $p,$ can be "compressed" till the length $-
\log
 p(x_1 ... x_t)$ bits and, on the other hand, there is no code $\psi$ for which the average codeword
 length (\,$  \Sigma_{x_1 ... x_t \in A^t} \, p(x_1 ... x_t) |\psi(x_1 ... x_t)|\,$ )
 is less than  $ - \Sigma_{x_1 ... x_t \in A^t} \, p(x_1 ... x_t)\log
 p(x_1 ... x_t)$.
 The universal codes can reach the lower bound
 $-
\log
 p(x_1 ... x_t)$ asymptotically for any stationary and ergodic source $p$ with  probability 1.
  The formal
definition is as follows: A code $\varphi$ is universal if for any
stationary and ergodic source $p$
\begin{equation}\label{un} \lim_{t \rightarrow \infty} t^{-1} (- \log
 p(x_1 ... x_t) - |\varphi(x_1 ... x_t)| ) \, = \,0 \end{equation}
 with  probability 1. So, informally speaking, universal
 codes estimate the probability characteristics of the source $p$ and
 use them for efficient "compression".
 One of the first universal
 codes was described in \cite{Ry0}, see also \cite{Ry1}. Now
 there are many efficient universal codes (and universal
 predictors connected with them), which are described in numerous papers, see \cite{S,Ki,
 No,Ri,Ry2}.

 \section{Identity Testing.}

 Now we consider the problem of testing $H_0^{id}$ against $H_1^{id}.$
  Let the required  level of significance (or a Type I
error) be $\alpha ,\, \alpha \in (0,1).$ (By definition, the Type
I error occurs if $H_0$ is true, but the test rejects $H_0$). We
describe a statistical test which can be constructed based on any
code $\varphi$.

The  main idea of the suggested test is quite natural: compress a
sample sequence $x_1... x_n$ by a code $\varphi$. If the length of
codeword ($|\varphi(x_1... x_n)|$) is significantly less than the
value $- \log \pi(x_1... x_n),$ then $H_0^{id}$ should be
rejected. The main observation is that the probability of all
rejected sequences is quite small for any $\varphi$, that is why
the Type I error can be made small. The precise description of the
test is as follows: { \it
 The hypothesis
$H_0^{id}$  is accepted if
\begin{equation}\label{t1} - \log
\pi(x_1... x_n) - |\varphi (x_1... x_n) | \leq - \log \alpha
.\end{equation}
  Otherwise, $H_0^{id}$ is rejected.}
We denote this test by $\Gamma_{\pi, \alpha,\varphi}^{(n)}.$

\textbf{Theorem 1.} { \it

i) For each distribution  $\pi, \alpha \in (0,1)$ and  a code
$\varphi$, the Type I error of the described test $\Gamma_{\pi,
\alpha,\varphi}^{(n)}$ is not larger than $\alpha .$

ii) If, in addition, $\pi$ is a finite-memory  stationary and
ergodic process  (i.e. $\pi \in M^*(A)$) and $\varphi$ is a
universal code, then the Type II error of the test $\Gamma_{\pi,
\alpha,\varphi}^{(n)}$ goes to 0, when $n$ tends to infinity.}

\textbf{Remark.} { \it The Kolmogorov complexity can be used
instead of the length of a code. Namely, let $K_{\pi,
\alpha}^{(n)}$ be the following test: the hypothesis $H_0^{id}$ is
accepted if $- \log \pi(x_1... x_n) - K(x_1... x_n)  \leq - \log
\alpha ,$ otherwise, $H_0^{id}$ is rejected. Theorem 1 is valid
for this test, too.}

\section{ Testing of Serial
Independence }

We first give some additional definitions. Let $v$ be a  word $v =
v_1 ... v_k, k \leq t, v_i \in A .$ Denote the rate of a word $v$
occurring in the sequence $x_1 x_2 \ldots x_k,$ $x_2x_3 \ldots
x_{k+1}$, $x_3x_4 \ldots x_{k+2}$, $ \ldots $, $x_{t-k+1} \ldots
x_t$ as $\nu^t(v)$. For example, if $x_1 ... x_t = 000100$ and $v=
00, $ then $ \nu^6(00) = 3$. Now we define for any $k \geq 0$ the
so-called empirical  Shannon entropy of  order $k$ as follows:
\begin{equation}\label{He}
h^*_{ k}( x_1 \ldots x_t) = -  \frac{1}{(t-k)}\sum_{v \in A^k}
\bar{\nu}^t(v)  \sum_{a \in A} (\nu^t(va) / \bar{\nu}^t(v)) \log
(\nu^t(va) / \bar{\nu}^t(v))\, ,
\end{equation} where $k < t$ and $ \bar{\nu}^t(v  )= \sum_{a \in A} \nu^t(v a ). $
In particular, if $k=0$, we obtain $ h^*_{ 0}( x_1 \ldots x_t) = -
\frac{1}{t} \sum_{a \in A} \nu^t(a)  \log (\nu^t(a) / t )\, , $

Let, as before, $H_0^{ind}$ be that the source $\pi$ is Markovian
with memory (or connectivity) not grater than $m,\: (m \geq 0),$
and the alternative hypothesis $H_1^{ind}$ be that the sequence is
generated by a stationary and ergodic source, which differs from
the source under $H_0^{ind}$. The suggested test is as follows.

\emph{Let $\psi$ be any code.
 By definition, the hypothesis $H_0^{ind}$ is accepted if
\begin{equation}\label{cr}
 (t-m) h^*_{m}(x_1 ... x_t) -  |\psi(x_1 ... x_t)|  \leq
\log (1 / \alpha) \,,
\end{equation} where $\alpha \in (0,1) .$  Otherwise, $H_0^{ind}$ is rejected.}
 We denote this
test by $\Upsilon_{\alpha,\,\psi, m}^{t}.$

\textbf{Theorem 2.} i) For any distribution $\pi$ and any code
$\psi$ the First Type error of the test $\Upsilon_{\alpha,\,\psi,
m}^{t}$ is less than or equal to $\alpha, \alpha \in (0,1)$.

ii) If, in addition, $\pi$ is a   stationary and ergodic process
over $A^\infty$  and $\psi$ is a universal code, then the Type II
error of the test $\Upsilon_{\alpha,\,\psi, m}^{t}$ goes to 0,
when $t$ tends to infinity.

\textbf{Comment.} If we use  Kolmogorov complexity $K(x_1... x_n)$
instead of the length of the code $|\psi(x_1 ... x_t)|, $ the
obtained test will have the same properties.

\section{ Experiments }

We applied the described method  of identity testing to
pseudorandom number generators. More precisely, we denote by $U$ a
source, which generates equiprobable and independent symbols from
the alphabet $\{0,1\}$ and consider the hypothesis $H_0^{id}$ that
a sequence is generated by $U$.

 We have taken
 linear
congruent generators (LCG), which are defined by the following
equality $$X_{n+1}=(A*X_n+C)\: mod\, M ,$$ where $X_{n}$ is the
$n$-th generated number \cite{K}. Each such  generator we will
denote by $LCG(M,A,C,X_0),$  where $X_0$ is the initial value of
the generator. Such generators are well studied and many of them
are used in practice, see  \cite{K}.

In our experiments we extract an eight-bit word from each
generated $X_i$ using the following algorithm. Firstly, the number
$\mu = \lfloor M/256 \rfloor $ was calculated and then each $X_i$
was transformed into an 8-bit word $\hat{X}_i$ as follows:

\begin{equation}\label{tr} \left.
\begin{array}{cc}
\hat{X}_i = \lfloor X_i/256 \rfloor\,\; if X_i < 256 \mu
\\
\hat{X}_i = empty\; word \,\; if X_i \geq 256 \mu
\end{array}
\right\} \end{equation}
 Then a sequence was compressed by the archiver \emph{ACE  v
1.2b} (see http://www.winace.com/). Experimental data about
testing of four linear congruent generators is given in the table.

\begin{table}[h]

\caption{Results of experiments}
\begin{tabular}{|c|c|c|}
\hline
%language and cipher/
$\;\;\qquad\qquad$ parameters $\qquad\qquad$ / length (bits) &400
000& 8 000 000
\\ M,A,C, $ X_0 $ & $ \quad $  & $ \quad $
\\ $ 10^8 +1,23,0,47594118 \qquad\quad$ & 390 240    & 7635936
\\ $ 2^{31},2^{16}+3,0,1 \quad \quad$ & extended & 7797984
\\ $ 2^{32},134775813,1,0 \quad \quad$ & extended & extended
\\$ 2^{32},69069,0,1 \quad \quad$ & extended & extended

\\
\hline
\end{tabular}
\end{table}

So, we can see from the first line of the table that the
$400000-$bit sequence generated by the LCG($ 10^8 +1,23,0,47594118
$) and transformed according to (\ref{tr}), was compressed to a
$390 240-$bit sequence. (Here  400000 is the length of the
sequence after transformation.) If we take the level of
significance $ \alpha \geq 2^{- 9760} $ and apply the test
$\Gamma_{U, \alpha,\varphi}^{(400000)}$,($\varphi = \emph{ACE  v
1.2b}$),  the hypothesis $H_0$ should be rejected, see Theorem 1
and (\ref{t1}). Analogously, the second line of the table shows
that the $8000000-$bit sequence generated by LCG($
2^{31},2^{16}+3,0,1 $) cannot be considered as random. ($H_0^{id}$
should be rejected if the level of significance $\alpha$ is
greater than $2^{- 202016} $.) On the other hand, the suggested
test accepts $H_0^{id}$ for the sequences generated by the two
latter generators, because the lengths of the ``compressed''
sequences increased.

The obtained information corresponds to the known data about the
generators mentioned above. Thus, it is shown in \cite{K} that the
first two generators are bad whereas the last two generators were
investigated in \cite{mo} and \cite{M1}, correspondingly, and are
regarded as good. So, we can see that the suggested testing is
quite efficient.

\section{Appendix.}
        The following well known inequality, whose proof can be found in
\cite{Ga}, will be used in proofs of both theorems.

 \textbf{Lemma.} Let $p$ and $q$ be
two probability distributions over some alphabet $B$. Then
$\sum_{b \in B} p(b) \log (p(b)/q(b) ) \geq 0$ with equality if
and only if $p=q.$

\textit{Proof} of Theorem 1.  Let $C_\alpha$ be a critical set of
the test $\Gamma_{\pi, \alpha,\varphi}^{(n)}$, i.e., by
definition, $ C_\alpha = \{ u: u \in A^t \,\,\& \: - \log \pi(u) -
|\varphi (u) | > - \log \alpha  \}. $ Let $\mu_\varphi$ be a
measure for which the claim 1 is true. We define an axillary set
$$ \hat{C}_\alpha = \{ u:  - \log \pi(u) - (- \log \mu_\varphi(u)
) > - \log \alpha  \}.
$$ We have
$$ 1 \geq \sum_{u \in \hat{C}_\alpha} \mu_\varphi(u)  \geq \sum_{u \in \hat{C}_\alpha}
\pi(u) / \alpha = (1/ \alpha) \pi(\hat{C}_\alpha) .$$ (Here the
second inequality follows from the definition of $\hat{C}_\alpha,$
whereas all others are obvious.) So, we obtain that
$\pi(\hat{C}_\alpha) \leq \alpha .$ From  definitions of
$C_\alpha, \hat{C}_\alpha $ and (\ref{kra}) we immediately obtain
that $\hat{C}_\alpha \supset C_\alpha.$ Thus, $\pi(C_\alpha) \leq
\alpha .$ By definition, $\pi(C_\alpha)$ is the value of the Type
I error. The first statement of the theorem 1 is proven.

Let us prove the second statement of the theorem. Suppose that the
hypothesis $H_1^{id}$ is true. That is, the sequence $x_1 \ldots
x_t$ is generated by some stationary and ergodic source $\tau$ and
$\tau \neq \pi .$ Our strategy is to show that
\begin{equation}\label{inf}    \lim _{t\rightarrow\infty}- \log \pi(x_1 \ldots
x_t) - |\varphi(x_1 \ldots x_t)| = \infty
\end{equation}
 with probability 1 (according to the
measure $\tau$). First we represent (\ref{inf}) as
\begin{multline*}
 - \log \pi(x_1
\ldots x_t) - |\varphi(x_1 \ldots x_t)| \\
= t ( \frac{1}{t} \log
\frac{\tau(x_1 \ldots x_t)}{\pi(x_1 \ldots x_t)}  + \frac{1}{t}( -
\log \tau(x_1 \ldots x_t) - |\varphi(x_1 \ldots x_t)| ) ).
\end{multline*}
 From
this equality and the property of a universal code (\ref{un}) we
obtain
\begin{equation}\label{inf2}  - \log \pi(x_1 \ldots
x_t) - |\varphi(x_1 \ldots x_t)| = t\, ( \frac{1}{t} \log
\frac{\tau(x_1 \ldots x_t)}{\pi(x_1 \ldots x_t)}  + o(1)).
\end{equation}
Now we use some results of the ergodic theory and the information
theory, which can be found, for ex., in \cite{Billingsley}.
Firstly, according to the Shannon-MacMillan-Breiman theorem, there
exists the limit $\lim_{t\rightarrow\infty} -  \log \tau(x_1
\ldots x_t) /t$ (with probability 1) and this limit is equal to
the so-called limit Shannon entropy, which we denote as
$h_\infty(\tau)$. Secondly, it is known that for any integer $k$
the following inequality is true: $h_\infty(\tau) \leq - \sum_{v
\in A^k} \tau(v) \sum_{a \in A} \tau(a/v) \log \tau(a/v).$ (Here
the right hand value is called $m-$ order conditional  entropy).
It will be convenient to represent both statements as follows:
\begin{equation}\label{inf3} \lim_{t\rightarrow\infty} -  \log \tau(x_1 \ldots x_t)
/t   \leq - \sum_{v \in A^k} \tau(v) \sum_{a \in A} \tau(a/v) \log
\tau(a/v)
\end{equation} for any $k \geq 0$ (with probability 1).
It is supposed that the process $\pi$ has a finite memory, i.e.
belongs to $M_s(A)$ for some $s$. Having taken into account the
definition of $M_s(A)$ (\ref{ma}), we obtain the following
representation:
\begin{multline*} - \log \pi(x_1 \ldots x_t)/t = - t^{-1} \sum_{i=1}^t \log
\pi(x_i/x_1 \ldots x_{i-1}) \\ = - t^{-1} (\sum_{i=1}^k \log
\pi(x_i/x_1 \ldots x_{i-1}) + \sum_{i=k+1}^t \log \pi(x_i/x_{i-k}
\ldots x_{i-1})) \end{multline*} for any $k \geq s.$  According to
the ergodic theorem there exists a limit
$$ \lim_{t\rightarrow\infty} t^{-1} \sum_{i=k+1}^t \log
\pi(x_i/x_{i-k} \ldots x_{i-1}),$$
 which  is equal to $ - \sum_{v
\in A^k} \tau(v) \sum_{a \in A} \tau(a/v) \log \pi(a/v),$ see
\cite{Billingsley, Ga}. So, from the two latter equalities we can
see that
$$  \lim_{t\rightarrow\infty} (- \log \pi(x_1 \ldots x_t))/t = - \sum_{v \in
A^k} \tau(v) \sum_{a \in A} \tau(a/v) \log \pi(a/v).$$ Taking into
account this equality, (\ref{inf3}) and (\ref{inf2}), we can see
that $$- \log \pi(x_1 \ldots x_t) - |\varphi(x_1 \ldots x_t)| \geq
t \,(\sum_{v \in A^k} \tau(v) \sum_{a \in A} \tau(a/v) \log
(\tau(a/v) / \pi(a/v) )) + o(t) $$ for any $k \geq s.$
   From this
inequality and the Lemma we can obtain that $ \,\,\,\,\,\,$ $-
\log \pi(x_1 \ldots x_t) - |\varphi(x_1 \ldots x_t)| \geq c\: t +
o(t)$, where $c$ is a positive constant, $t\rightarrow\infty.$
Hence, (\ref{inf}) is true and the theorem is proven.

\textit{Proof} of Theorem 2. First we show that for any source
$\theta^* \in M_0(A)$ and any word $x_1 \ldots x_t \in A^t, t > 1,
$ the following inequality is valid:
\begin{equation}\label{ta}
\theta^* (x_1 \ldots x_t) = \prod_{a \in A}
(\theta^*(a))^{\nu^t(a)} \leq \prod_{a \in A} (
\nu^t(a)/t)^{\nu^t(a)} \end{equation} Here the equality holds,
because $\theta^* \in M_0(A)$ . The inequality follows from the
Lemma. Indeed, if $ p(a) = \nu^t(a)/t$ and $ q(a) = \theta^* (a),$
then $ \sum_{a \in A} \frac{\nu^t(a)}{t} \log
\frac{(\nu^t(a)/t)}{\theta^*(a) }\geq 0. $ From the latter
inequality we obtain (\ref{ta}).

Let now $\theta$ belong to $M_m(A), m > 0.$ We will prove that for
any $x_1 \ldots x_t $
\begin{equation}\label{taa}
\theta (x_1 \ldots x_t) \leq \prod_{u \in A^m } \prod_{a \in A} (
\nu^t(ua)/\bar{\nu}^t(u))^{\nu^t(ua)} \:. \end{equation}
Indeed,
we can present $\theta (x_1 \ldots x_t)$ as
$$\theta (x_1 \ldots
x_t)= \theta(x_1 \ldots x_m) \prod_{u \in A^m } \prod_{a \in A}
\theta (a/u)^{\nu^t(ua)}\:,$$
where $\theta(x_1 \ldots x_m)$
 is
the limit probability of the word $x_1 \ldots x_m .$ Hence,
$\theta (x_1 \ldots x_t) \leq \prod_{u \in A^m } \prod_{a \in A}
\theta (a/u)^{\nu^t(ua)}\:.$ Taking into account the inequality
(\ref{ta}), we obtain
$$\prod_{a \in A}
\theta (a/u)^{\nu^t(ua)} \leq  \prod_{a \in A} (
\nu^t(ua)/\bar{\nu}^t(u))^{\nu^t(ua)}$$
 for any word $u$. So, from
the last two  inequalities we obtain (\ref{taa}).

It will be  convenient to define two auxiliary measures  on $A^t$
as follows:
\begin{equation}\label{ro}
\pi_m (x_1 ... x_t) = \Delta \:2^{- t\, h^*_m(x_1 ... x_t)}\,,
\:\sigma(x_1 ... x_t) = 2^{-|\psi(x_1 ... x_t)|}
\end{equation} where $x_1 ... x_t \in A^t $ and
$\Delta = (\sum_{x_1 ... x_t \in A^t}\,2^{- t \,h^*_m (x_1 ...
x_t)}\,)^{- 1}\,.$ If we take into account that $2^{- (t-m) \,
h^*_m(x_1 ... x_t)}\, = \prod_{u \in A^m } \prod_{a \in A} (
\nu^t(ua)/\bar{\nu}^t(u))^{\nu^t(ua)} \:,$ we can see from
(\ref{taa}) and (\ref{ro}) that, for any  measure $\theta \in
M_m(A)$ and any $x_1 \ldots x_t \in A^t,   $
\begin{equation}\label{r11} \theta (x_1 \ldots x_t) \leq \pi_m  (x_1 ... x_t) / \Delta \:.
\end{equation}
Let us denote the critical set of the test
$\Upsilon_{\alpha,\,\sigma, m}^{t} $ as $C_\alpha,$ i.e., by
definition, $ C_\alpha = \{ x_1 \ldots x_t :\; (t - m)\: h^*_m(x_1
\ldots x_t) - |\psi(x_1 ... x_t)| )
>
\log (1 / \alpha)   \} . $ From (\ref{ro}) we obtain
\begin{equation}\label{C} C_\alpha = \{ x_1 \ldots x_t :\; (t - m)\: h^*_m(x_1 \ldots x_t) -
(- \log \sigma(x_1 ... x_t)) \,)
>
\log (1 / \alpha)   \} . \end{equation}
    From (\ref{r11}) and
(\ref{C}) we can see that for any measure $\theta \in M_m(A)$
\begin{equation}\label{r1}
\theta (C_\alpha) \leq \pi_m  (C_\alpha) / \Delta \:.
\end{equation}
From (\ref{C}) and (\ref{ro}) we obtain
\begin{multline*} C_\alpha = \{ x_1 \ldots x_t :\;  2^{\,(t-m)\: h^*_m(x_1 \ldots x_t)} >
(\alpha \,\:\sigma(x_1 \ldots x_t) )^{-1}  \} \\ = \{ x_1 \ldots
x_t :\; ( \pi_m(x_1 \ldots x_t ) / \Delta )^{-1}   > (\alpha
\,\:\sigma(x_1 \ldots x_t) )^{-1} \} \:.        \end{multline*}
Finally,
\begin{equation}\label{C1} C_\alpha = \{ x_1 \ldots x_t :\; \sigma(x_1 \ldots
x_t) > \pi_m(x_1 \ldots x_t ) / (\alpha\, \Delta ) \} .
\end{equation}
The following chain of inequalities and equalities is valid:
\begin{multline*} 1 \geq \sum_{x_1 \ldots x_t \in C_\alpha} \sigma( x_1 \ldots
x_t) \geq \sum_{x_1 \ldots x_t \in C_\alpha} \pi_m(x_1 \ldots x_t
) / (\alpha\, \Delta )\\ = \pi_m(C_\alpha)/ (\alpha\, \Delta )
\geq \theta(C_\alpha) \Delta / (\alpha\, \Delta ) =
\theta(C_\alpha)
 / \alpha .\end{multline*}
  (Here both equalities and the first inequality are
 obvious, the second   and the third inequalities follow from
(\ref{C1}) and (\ref{r1}), correspondingly.) So, we obtain that
$\theta(C_\alpha) \leq \alpha $ for any measure $\theta \in
M_m(A).$ Taking into account that $C_\alpha$ is the critical set
of the test, we can see that the probability of the First Type
error is not greater than $\alpha.$ The first claim of the theorem
is proven.

The proof of the second  statement of the theorem  will  be based
on some results of Information Theory. The $t-$ order conditional
Shannon entropy is defined as follows:
 \begin{equation}\label{SHE} h_t(p) =   -
\sum_{x_1 ... x_t \in A^t} p(x_1 ... x_t) \sum_{a \in A} p(a/x_1
... x_t) \log p(a/x_1 ... x_t),
\end{equation} where $p \in M_\infty(A).$ It is known that for any
$p \in M_\infty(A)$ firstly, $\log |A| \geq h_0(p) \geq h_1(p)
\geq ... ,$ secondly, there exists  limit Shannon entropy
$h_\infty (p) = \lim_{t\rightarrow\infty} h_t(p) $,  thirdly,
$\lim_{t\rightarrow\infty} - t^{-1} \log p(x_1 ... x_t) = h_\infty
(p) $ with  probability 1 and, finally,
 $h_m(p)$ is strictly greater than $h_\infty(p), $ if
the memory of $p$ is grater than $m$, (i.e. $p \in M_\infty(A)
\setminus M_m(A)$), see, for example, \cite{Billingsley, Ga}.

Taking into account the definition of the universal code
(\ref{un}), we obtain from the above described properties of the
entropy that
\begin{equation}\label{unh}
\lim_{t\rightarrow\infty}  t^{-1} |\psi(x_1 ... x_t)| = h_\infty
(p)
\end{equation} with probability 1.
    It can be seen from (\ref{He}) that $h^*_m$
 is an estimate for the $m-$order Shannon
entropy  (\ref{SHE}). Applying the ergodic theorem we obtain $
\qquad\qquad\lim_{t\rightarrow\infty} h^*_m (x_1 \ldots x_t ) =
h_m(p)$ with probability 1; see \cite{Billingsley, Ga}. Having
taken into account that $h_m(p) > h_\infty (p)$ and (\ref{unh}) we
obtain from the last equality that $\lim_{t\rightarrow\infty} ( (t
- m) \,h^*_m (x_1 \ldots x_t )  - |\psi(x_1 ... x_t)|) = \infty .$
This proves the second statement of the theorem.

%\newpage

%**********************************************************
%*                   Bibliography                         *
%**********************************************************
%\newpage

\end{document}